\documentclass[a4paper]{jpconf}
\usepackage{graphicx}
\usepackage{color}
\usepackage{amsmath,amssymb}

\begin{document}
\title{Testing the hadronic spectrum in the strange sector}
\author{Paolo Parotto}
\address{Department of Physics, University of Houston, Houston, TX 77204, USA}
\ead{pparotto@uh.edu}

\begin{abstract}
Heavier resonances are continually being added to the hadronic spectrum from the Particle Data Group that follow an exponentially increasing mass spectrum.  However, it has been suggested that even further states predicted from Quark Models are needed in the hadronic spectrum in order to improve the agreement between the hadron resonance gas model predictions and lattice QCD data. We find that the inclusion of such states with extrapolated branching ratios slightly decreases the freezeout temperature. To eliminate ambiguities, we introduce a first principle method to extract the freeze-out temperature for charged kaons from experimental data, which yields a lower bound of $T_{\text{fo}} \gtrsim $145 MeV for the highest collision energy at RHIC.
\end{abstract}

\section{Introduction}
The Hadron Resonance Gas model has been extremely successful in reproducing Lattice Quantum Chromodynamics results in the hadron gas phase \cite{Ratti:2010kj,Bluhm:2013yga}. The underlying assumption is that it is possible to describe an interacting gas of hadrons in the ground states through a non-interacting gas of hadrons and resonances. The expression for thermodynamic quantities in the HRG model is then simply a sum over the contributions from each resonance. In this sense, the only ``variable'' in the model is the hadronic spectrum one feeds into it. 

The original idea by Rolf Hagedorn was that a limiting temperature $T_H$ for matter exists, called Hagedorn temperature, which cannot be exceeded, and therefore a system of hadrons responds to an arbitrary increase in the energy density by creating more and more resonances rather than by increasing the temperature \cite{Hagedorn:1965st}. Following these arguments, he showed that these resonances would need to populate a spectrum which is exponentially increasing in mass, when counting all states with the corresponding degeneracy:
$$N(m) = \sum_i  d_i \, \Theta_i(m-m_i) $$
summing over the resonances with degeneracy $d_i$ and mass $m_i$.  It was later shown  \cite{Cabibbo:1975ig} that $T_H$ can be interpreted as the limiting temperature for hadronic matter, if a phase transition occurs to a deconfined state of quarks and gluons.  Presently, we know from Lattice QCD calculations that the deconfinement transition is a smooth crossover for vanishing baryonic chemical potential \cite{Aoki:2006we}. In this sense, the Hagedorn temperature can be  analogous to the critical temperature of the deconfinement transition, and, therefore, it is reasonable to expect the effects related to the increased states population to appear and become relevant close to such transition.  

\begin{figure}[t]
\begin{center}
\includegraphics[width=0.6\textwidth]{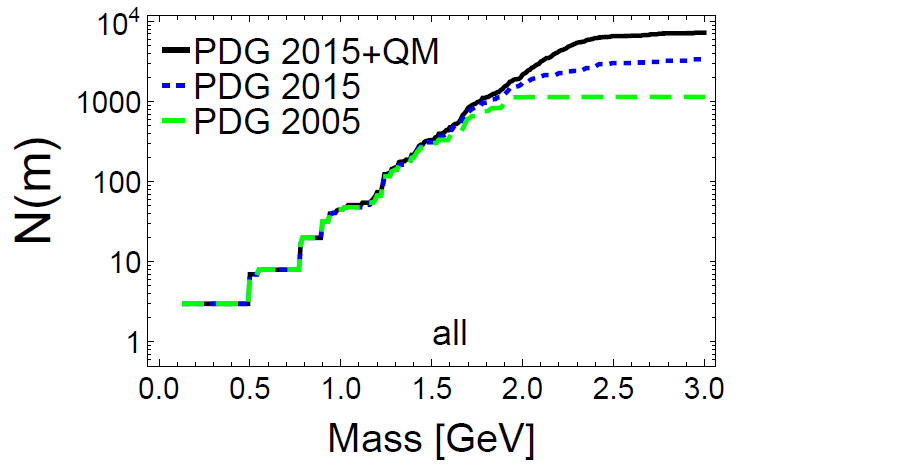}
\caption{Exponentially increasing mass spectrum for different PDG lists, and with the inclusion of further Quark Model states} \label{HAG_spec}
\end{center}
\end{figure}

The list of experimentally measured resonances is continuously updated by the Particle Data Group \cite{Agashe:2014kda}, and confirms the original idea from Hagedorn  \cite{Broniowski:2004yh,Lo:2015cca}. Moreover, there exist Quark Models \cite{Capstick:1986bm,Godfrey:1985xj} that predict the presence of further, not yet discovered, hadronic states. The latter states also show an exponential behavior in the mass spectrum, as shown in Fig.\ref{HAG_spec}. Including further states in an exponentially increasing mass spectrum can have far reaching effects in the field of heavy-ions.  For instance, they play a role in the transport coefficients close to the critical temperature region  \cite{NoronhaHostler:2008ju,NoronhaHostler:2012ug,Kadam:2014cua,Pal:2010es}, influence elliptical flow \cite{Noronha-Hostler:2013rcw,Paquet:2015lta},   can improve thermal fits \cite{NoronhaHostler:2009tz}, and affect the chemical equilibration time in the hadron gas phase \cite{NoronhaHostler:2007jf,NoronhaHostler:2009cf,Beitel:2014kza,Beitel:2016ghw}. Thus, it is important to determine the precise number of resonances that exist otherwise a systematic bias affects numerous comparisons between theory and experiment. 

\section{Strange states and chemical freeze-out}

The Hadron Resonance Gas model has been employed to extract information about chemical freeze-out in heavy ion collisions through comparison of particle yields \cite{Stachel:2013zma}, particle ratios \cite{Andronic:2008gu} and ratios of fluctuations of conserved charges (susceptibilities) \cite{Alba:2014eba}, with experimental data. In the past few years, such comparisons of yields and particle ratios with data from both RHIC and the LHC seemed to indicate the existence of different freeze-out temperatures for light and strange matter \cite{Floris:2014pta}. When comparing to the experiment, effects due to resonance decays, acceptance cuts and isospin randomization need to be taken into account \cite{Alba:2014eba}, which can also be  affected by the inclusion of more states in the hadronic population  \cite{Noronha-Hostler:2014usa,Noronha-Hostler:2014aia}.  

\begin{figure}
  \centering
  \begin{minipage}[b]{0.45\textwidth}
\includegraphics[width=\textwidth]{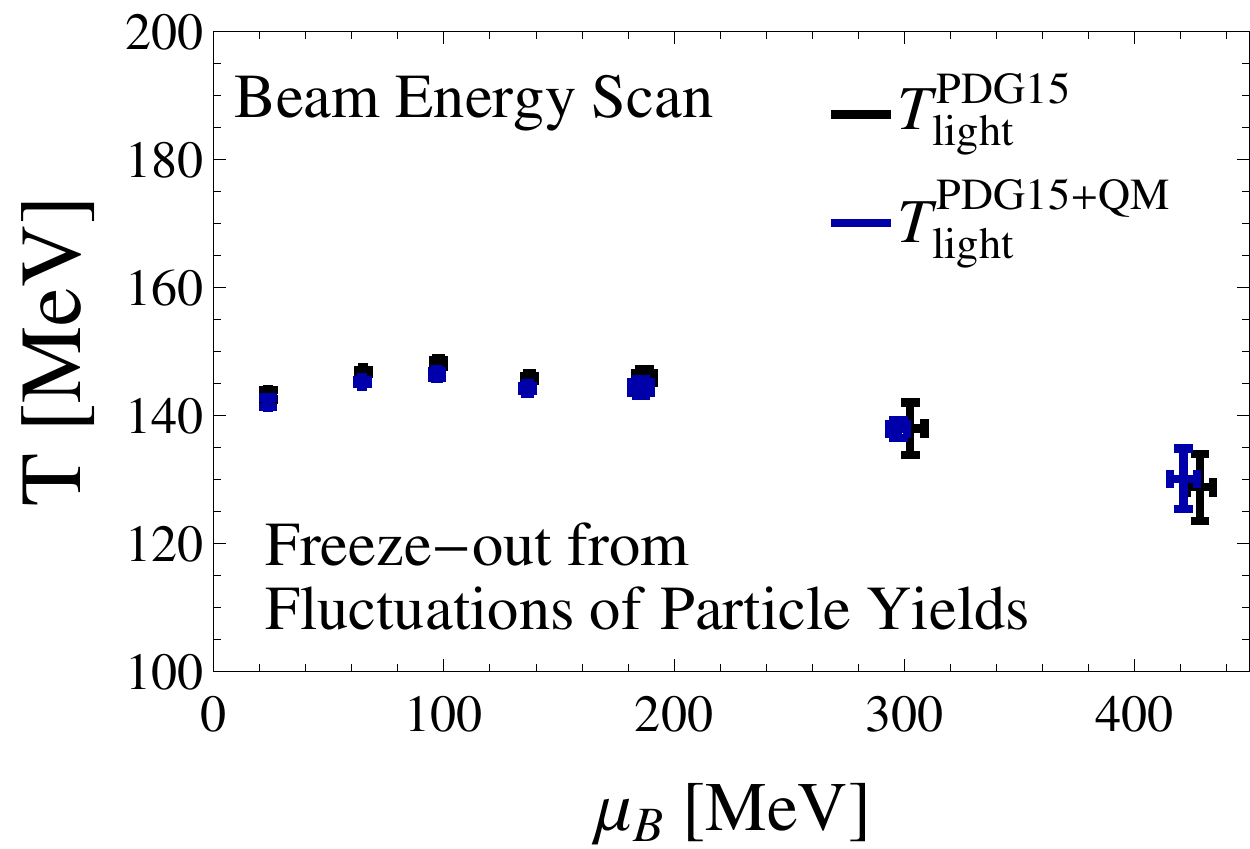}
\caption{Freeze-out points from particle yields fluctuations comparison with HRG with and without Quark Models states} \label{Tmub}
\vspace{4mm}
  \end{minipage}
  \hfill
  \begin{minipage}[b]{0.45\textwidth}
\includegraphics[width=\textwidth]{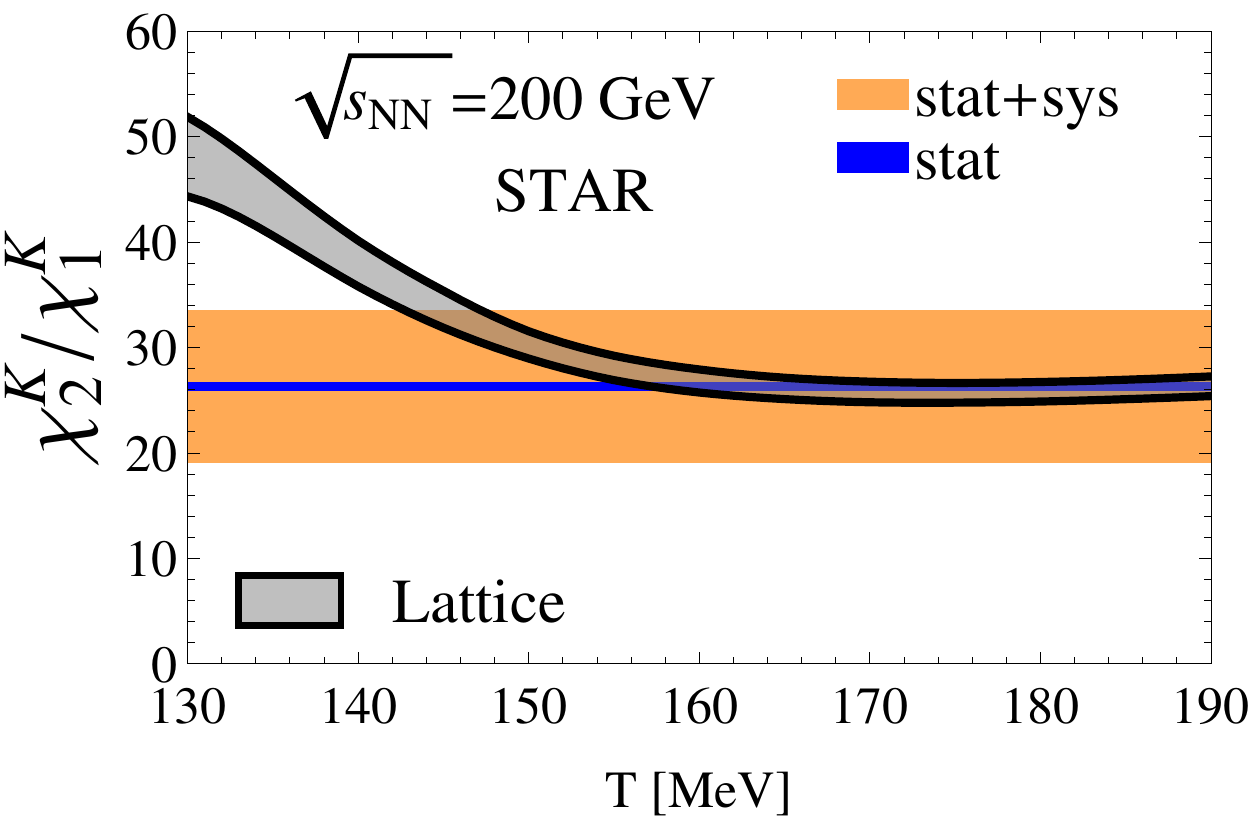}
\caption{Comparison of charged kaons $\chi^K_2/\chi^K_1$ from Lattice QCD (Wuppertal Budapest) to preliminary STAR data for $\sqrt{s_{NN}}=$200 GeV.} \label{Lat_STAR} 
  \end{minipage}
\end{figure}

For strangeness, the agreement of the Hadron Resonance Gas model with Lattice results  is not always satisfactory, so that it was suggested \cite{Bazavov:2014xya} that the inclusion of Quark model states is necessary. It was also pointed out that this inclusion would reduce the difference in temperature between light and strange freeze-out. However, in \cite{Bazavov:2014xya} the effects of resonance decays were not taken into account. Thus, we used the known branching ratios from the PDG list \cite{Agashe:2014kda} to extrapolate the branching ratios for the Quark Model states, carefully enforcing quantum numbers and mass conservation. By comparing net-charge and net-proton fluctuations ($\chi_1/\chi_2$) with STAR data \cite{Luo:2015ewa,Adamczyk:2014fia} from the Beam Energy Scan, it was possible to extract the values of $T$ and $\mu_B$ at chemical freeze-out. The addition of Quark model states only has a small decrease of the freeze-out temperature, as seen in Fig.\ref{Tmub}. \\
To extract freeze-out parameters from first principles, one can use ratios of susceptibilities as discussed in \cite{Karsch:2012wm,Borsanyi:2014ewa}. If Lattice QCD showed a separation between light and strange freeze-out \cite{Bellwied:2013cta}, this would have an heavy impact for example on the hadronization schemes used in hydrodynamical simulations. In principle, one could compare Lattice QCD to experiment directly from susceptibilities of strangeness.  However, lattice  only has global contributions whereas the only strangeness-related fluctuations presently available from experiment are those of charged kaons. In order to isolate charged kaons on the lattice, partial pressures for charged strange mesons can be used \cite{Noronha-Hostler:2016rpd}, where ratios of susceptibilities are shown to be of the form 
\begin{equation}\label{fluc_ratio}
\frac{\chi^K_2}{\chi^K_1} = \frac{\cosh(\hat{\mu}_S+\hat{\mu}_Q)}{\sinh(\hat{\mu}_S+\hat{\mu}_Q)} \, \, ,
\end{equation}
where $\hat{\mu}_S=\mu_S/T$ and $\hat{\mu}_Q=\mu_Q/T$ are supplied by the Wuppertal Budapest collaboration. \\
In Fig.\ref{Lat_STAR} $\frac{\chi^K_2}{\chi^K_1}$ is compared from Lattice QCD with preliminary STAR data. Due to large uncertainties in the experimental data, only a lower bound at $T^K_{fo} \gtrsim 145 \, \text{MeV}$ can be extracted at the moment.

\section{Conclusions}
In this proceedings two different approaches for studying  light and strange freeze-out temperatures are shown. Adding Quark model states with their decays to the hadronic spectrum, lowers the light freeze-out temperature by a small amount. Second, a new method for extracting strangeness freeze-out parameters from first principles using strange charged sucpetibilities is presented to resolve the tension between the light and strange sector. In both cases, smaller error bars are needed to better check the strange sector.

\section*{Acknowledgments}
This material is based upon work supported by the National Science Foundation under grant
no. PHY-1513864 and by the U.S. Department of Energy, Office of Science, Office of Nuclear
Physics, within the framework of the Beam Energy Scan Theory (BEST) Topical Collaboration.
This work contains lattice QCD data provided by the Wuppertal-Budapest Collaboration. An
award of computer time was provided by the INCITE program. This research used resources
of the Argonne Leadership Computing Facility, which is a DOE Office of Science User Facility
supported under Contract DE-AC02-06CH11357. The work of J. G. and A. P. was supported
by the DFG grant SFB/TR55. The authors gratefully acknowledge the Gauss Centre for Supercomputing (GCS) for providing computing time for a GCS Large-Scale Project on the
GCS share of the supercomputer JUQUEEN [30] at J\"ulich Supercomputing Centre (JSC).


\section*{References}
\begin{thebibliography}{9}

\bibitem{Ratti:2010kj} 
  C.~Ratti {\it et al.} [Wuppertal-Budapest Collaboration],
  Nucl.\ Phys.\ A {\bf 855}, 253 (2011)
  doi:10.1016/j.nuclphysa.2011.02.052

\bibitem{Bluhm:2013yga} 
  M.~Bluhm, P.~Alba, W.~Alberico, A.~Beraudo and C.~Ratti,
  Nucl.\ Phys.\ A {\bf 929}, 157 (2014)
  doi:10.1016/j.nuclphysa.2014.06.013

\bibitem{Hagedorn:1965st} 
  R.~Hagedorn,
  Nuovo Cim.\ Suppl.\  {\bf 3}, 147 (1965).

\bibitem{Cabibbo:1975ig} 
  N.~Cabibbo and G.~Parisi,
  Phys.\ Lett.\ B {\bf 59}, 67 (1975).
  doi:10.1016/0370-2693(75)90158-6


\bibitem{Aoki:2006we} 
  Y.~Aoki, G.~Endrodi, Z.~Fodor, S.~D.~Katz and K.~K.~Szabo,
  Nature {\bf 443}, 675 (2006)
  doi:10.1038/nature05120
  
  
\bibitem{Agashe:2014kda} 
  K.~A.~Olive {\it et al.} [Particle Data Group Collaboration],
  Chin.\ Phys.\ C {\bf 38}, 090001 (2014).
  doi:10.1088/1674-1137/38/9/090001
  
 
\bibitem{Noronha-Hostler:2014aia} 
  J.~Noronha-Hostler and C.~Greiner,
  Nucl.\ Phys.\ A {\bf 931}, 1108 (2014)
  doi:10.1016/j.nuclphysa.2014.08.101

\bibitem{Noronha-Hostler:2014usa} 
  J.~Noronha-Hostler and C.~Greiner,
  arXiv:1405.7298 [nucl-th].
  
  
\bibitem{Broniowski:2004yh} 
  W.~Broniowski, W.~Florkowski and L.~Y.~Glozman,
  Phys.\ Rev.\ D {\bf 70}, 117503 (2004)
  doi:10.1103/PhysRevD.70.117503
  
\bibitem{Lo:2015cca} 
  P.~M.~Lo, M.~Marczenko, K.~Redlich and C.~Sasaki,
  Phys.\ Rev.\ C {\bf 92}, no. 5, 055206 (2015)
  doi:10.1103/PhysRevC.92.055206
  
\bibitem{Capstick:1986bm} 
  S.~Capstick and N.~Isgur,
  Phys.\ Rev.\ D {\bf 34}, 2809 (1986)
  [AIP Conf.\ Proc.\  {\bf 132}, 267 (1985)].
  doi:10.1103/PhysRevD.34.2809, 10.1063/1.35361
  
  
\bibitem{Godfrey:1985xj} 
  S.~Godfrey and N.~Isgur,
  Phys.\ Rev.\ D {\bf 32}, 189 (1985).
  doi:10.1103/PhysRevD.32.189
  
\bibitem{NoronhaHostler:2008ju} 
  J.~Noronha-Hostler, J.~Noronha and C.~Greiner,
  Phys.\ Rev.\ Lett.\  {\bf 103}, 172302 (2009)
  doi:10.1103/PhysRevLett.103.172302
\bibitem{NoronhaHostler:2012ug} 
  J.~Noronha-Hostler, J.~Noronha and C.~Greiner,
  Phys.\ Rev.\ C {\bf 86}, 024913 (2012)
  doi:10.1103/PhysRevC.86.024913
\bibitem{Kadam:2014cua} 
  G.~P.~Kadam and H.~Mishra,
  Nucl.\ Phys.\ A {\bf 934}, 133 (2014)
  doi:10.1016/j.nuclphysa.2014.12.004
\bibitem{Pal:2010es} 
  S.~Pal,
  Phys.\ Lett.\ B {\bf 684}, 211 (2010)
  doi:10.1016/j.physletb.2010.01.017
  
\bibitem{Noronha-Hostler:2013rcw} 
  J.~Noronha-Hostler, J.~Noronha, G.~S.~Denicol, R.~P.~G.~Andrade, F.~Grassi and C.~Greiner,
  Phys.\ Rev.\ C {\bf 89}, no. 5, 054904 (2014)
  doi:10.1103/PhysRevC.89.054904
\bibitem{Paquet:2015lta} 
  J.~F.~Paquet, C.~Shen, G.~S.~Denicol, M.~Luzum, B.~Schenke, S.~Jeon and C.~Gale,
  Phys.\ Rev.\ C {\bf 93}, no. 4, 044906 (2016)
  doi:10.1103/PhysRevC.93.044906
  
  

  
  
\bibitem{NoronhaHostler:2009tz} 
  J.~Noronha-Hostler, H.~Ahmad, J.~Noronha and C.~Greiner,
  Phys.\ Rev.\ C {\bf 82}, 024913 (2010)
  doi:10.1103/PhysRevC.82.024913
  
  
\bibitem{NoronhaHostler:2007jf} 
  J.~Noronha-Hostler, C.~Greiner and I.~A.~Shovkovy,
  Phys.\ Rev.\ Lett.\  {\bf 100}, 252301 (2008)
  doi:10.1103/PhysRevLett.100.252301
\bibitem{NoronhaHostler:2009cf} 
  J.~Noronha-Hostler, M.~Beitel, C.~Greiner and I.~Shovkovy,
  Phys.\ Rev.\ C {\bf 81}, 054909 (2010)
  doi:10.1103/PhysRevC.81.054909
\bibitem{Beitel:2014kza} 
  M.~Beitel, K.~Gallmeister and C.~Greiner,
  Phys.\ Rev.\ C {\bf 90}, no. 4, 045203 (2014)
  doi:10.1103/PhysRevC.90.045203
\bibitem{Beitel:2016ghw} 
  M.~Beitel, C.~Greiner and H.~Stoecker,
  Phys.\ Rev.\ C {\bf 94}, no. 2, 021902 (2016)
  doi:10.1103/PhysRevC.94.021902

\bibitem{Stachel:2013zma} 
  J.~Stachel, A.~Andronic, P.~Braun-Munzinger and K.~Redlich,
  J.\ Phys.\ Conf.\ Ser.\  {\bf 509}, 012019 (2014)
  doi:10.1088/1742-6596/509/1/012019

\bibitem{Andronic:2008gu} 
  A.~Andronic, P.~Braun-Munzinger and J.~Stachel,
  Phys.\ Lett.\ B {\bf 673}, 142 (2009)
  Erratum: [Phys.\ Lett.\ B {\bf 678}, 516 (2009)]
  doi:10.1016/j.physletb.2009.02.014, 10.1016/j.physletb.2009.06.021


\bibitem{Alba:2014eba} 
  P.~Alba, W.~Alberico, R.~Bellwied, M.~Bluhm, V.~Mantovani Sarti, M.~Nahrgang and C.~Ratti,
  Phys.\ Lett.\ B {\bf 738}, 305 (2014)
  doi:10.1016/j.physletb.2014.09.052


\bibitem{Floris:2014pta} 
  M.~Floris,
  Nucl.\ Phys.\ A {\bf 931}, 103 (2014)
  doi:10.1016/j.nuclphysa.2014.09.002




\bibitem{Noronha-Hostler:2016rpd} 
  J.~Noronha-Hostler, R.~Bellwied, J.~Gunther, P.~Parotto, A.~Pasztor, I.~P.~Vazquez and C.~Ratti,
  arXiv:1607.02527 [hep-ph].
  


 


 
\bibitem{Bazavov:2014xya} 
  A.~Bazavov {\it et al.},
  Phys.\ Rev.\ Lett.\  {\bf 113}, no. 7, 072001 (2014)
  doi:10.1103/PhysRevLett.113.072001



\bibitem{Luo:2015ewa} 
  X.~Luo [STAR Collaboration],
  PoS CPOD {\bf 2014}, 019 (2015)

\bibitem{Adamczyk:2014fia} 
  L.~Adamczyk {\it et al.} [STAR Collaboration],
  Phys.\ Rev.\ Lett.\  {\bf 113}, 092301 (2014)
  doi:10.1103/PhysRevLett.113.092301

  
\bibitem{Karsch:2012wm} 
  F.~Karsch,
  Central Eur.\ J.\ Phys.\  {\bf 10}, 1234 (2012)
  doi:10.2478/s11534-012-0074-3


\bibitem{Borsanyi:2014ewa} 
  S.~Borsanyi, Z.~Fodor, S.~D.~Katz, S.~Krieg, C.~Ratti and K.~K.~Szabo,
  Phys.\ Rev.\ Lett.\  {\bf 113}, 052301 (2014)
  doi:10.1103/PhysRevLett.113.052301

\bibitem{Bellwied:2013cta} 
  R.~Bellwied, S.~Borsanyi, Z.~Fodor, S.~D.~Katz and C.~Ratti,
  Phys.\ Rev.\ Lett.\  {\bf 111}, 202302 (2013)
  doi:10.1103/PhysRevLett.111.202302


  
\end{thebibliography}
\end{document}